# Measurement of two-point coherence functions of electromagnetic optical fields and applications of optical coherence

Bhaskar Kanseri[1] and Deepa Joshi

*Department of Physics, Indian Institute of Technology Delhi, Hauz Khas, New Delhi-110016, India*

For stationary light fields, manifestation of statistical properties such as coherence and polarization are attributed to the same physical phenomena, i.e. correlations in fluctuations of optical fields. In order to explain various properties associated with electromagnetic optical fields, both coherence and polarization need to be placed at same footings. This leads to two-point (space or time) generalization of single-point properties such as Stokes parameters and elements of coherency matrix. This paper reviews the basic aspects concerning vectorial optical fields and experimental methods developed during last couple of decades for the measurement of two-point correlation functions of electromagnetic optical fields in spatial and temporal domain. Studies related to coherence properties of optical fields have led to several important technological applications during last seven decades, which are also discussed briefly in this review.

## 1. Introduction

Since long ago, statistical features of light such as coherence and polarization were known to be intimately connected [1, 2] and the correlations between orthogonal electric field components of a partially polarized beam could not be satisfactorily explained by using only coherence and incoherence and consequently it was desired to invoke intermediate states namely partial coherence [3, 4]. Initial investigations of statistical properties of electromagnetic (EM) radiation in connection with partially coherent scalar light fields were first proposed [5]. The correlation between orthogonal electric fields components is governed by a 2x2 matrix namely the coherency (polarization) matrix proposed by E. Wolf in 1959 [4] and a systematic analysis of coherence properties of partially polarized radiation was first made. It was shown in this work that the maximum value of the degree of coherence between the orthogonal components of the light field having equal intensities is equal to the degree of polarization of the field. The Stokes parameters [6], which provide a mathematical formulation of polarization properties of light fields are shown to be simple linear combinations of the elements of coherency matrix and it was concluded that both the unique representations are essentially equivalent [7, 8]. Interestingly, both Stokes parameters and elements of coherency matrix turn out to be measurable quantities [5, 9, 10]. For scalar light fields, innumerable studies and developments have been made during last six decades including the legacy work by E. Wolf and coworkers, which include measurement of correlation functions for several kinds of beams, and their applications in wide areas of physics [11-14].

The partial coherence theory proposed for scalar optical fields works perfectly for light fields having no variation in the polarization state [12]. However, light fields may have partially coherent and partially polarized nature in which field components may contain different coherence features [15]. Such fields were characterized as electromagnetic (EM) fields, or vectorial optical fields [16]. It has been observed that coherence and polarization properties of such fields are unified together and cannot be treated independently [17]. Several new kinds of effects such as change in polarization of light on propagation etc. were explained treating light fields as electromagnetic [18-20]. Partial coherence has been investigated in connection with EM fields mainly by E. Wolf, AT. Friberg and other coworkers during past couple of decades [21]. One can have two-point correlations both in spatial and in temporal domain. Unlike in the scalar coherence theory, there was no single scalar quantity that was capable of describing the coherence of electromagnetic fields at two separate space–time points. Hence the correlation properties were initially examined by using the concept of the degree of polarization, which was capable of describing the correlations at a one point only [15]. Later it was realized that the measurement of coherence cannot be made using a single visibility parameter and one needs to take all cross-correlations also into account [22]. Such degree of coherence of EM fields has been thus proposed and parameters related to coherence-polarization matrices along with two-point correlation functions (spatial and temporal) have been measured in space-time and space-frequency domain during last couple of decades [23-42].

This article reviews some of the methods proposed and implemented for the determination of the two-point correlation functions of EM optical fields during last several decades. Beginning with a brief discussion on principles associated with electromagnetic optical fields, more focus is given to experimental methods used for the determination of these parameters using interferometric tools, such as consisting of Young's type double-slit interferometers (based on amplitude correlation) and based on intensity correlation based interferometer. Towards the end, wide ranging applications of optical coherence are discussed which cover imaging, interferometry, fundamental studies, quantum applications, optical communication etc.

---

[1] *Corresponding author*
*e-mail: bkanseri@physics.iitd.ac.in*

## 2. Two-point spatial correlation functions of EM optical fields

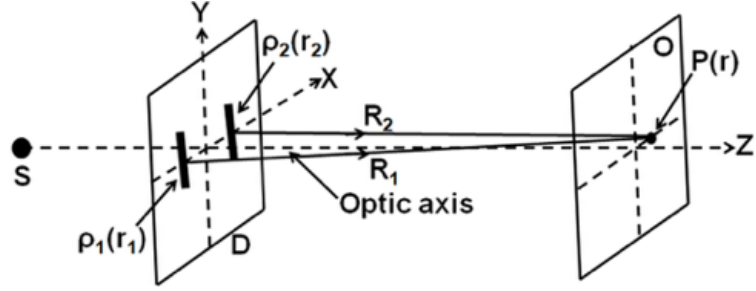

Fig 1. Schematic representation of Young's double-slit experiment. Notations: S secondary source, D double-slit, and O observation plane. P(**r**) is an off-axis point which is $R_1$ and $R_2$ distances away from the slits having position vectors **r**$_1$ and **r**$_2$, respectively.

Let us consider a random electromagnetic beam propagating in z-direction. The fluctuations of the beam could be considered statistically stationary at least in wide sense. Let $E_i(\mathbf{r}, \omega)$ for $(i = x, y)$, be the statistical ensemble of frequency $\omega$ of the fluctuating electric field $E(\mathbf{r}, \omega)$ at a point **r** in space. The second order coherence properties of the beam may be characterized by a 2×2 electric cross-spectral density (CSD) matrix [15]

$$\tilde{W}(\mathbf{r}_1,\mathbf{r}_2,\omega) = W_{ij}(\mathbf{r}_1,\mathbf{r}_2,\omega) = \langle E_i^*(\mathbf{r}_1,\omega)E_j(\mathbf{r}_2,\omega)\rangle, \; i = x, y, j = x, y \quad (1)$$

where the asterisk denotes the complex conjugate and angular brackets denote ensemble average. Eq. (1) represents the correlations in space-frequency domain and similar representation in space-time domain can be used in terms of a 2×2 beam coherence-polarization matrix (BCP) matrix [16, 43]

$$J(\mathbf{r}_1,\mathbf{r}_2) = \begin{bmatrix} J_{xx}(\mathbf{r}_1,\mathbf{r}_2) & J_{xy}(\mathbf{r}_1,\mathbf{r}_2) \\ J_{yx}(\mathbf{r}_1,\mathbf{r}_2) & J_{yy}(\mathbf{r}_1,\mathbf{r}_2) \end{bmatrix}, \quad (2)$$

where the elements of the matrix can be expressed in terms of the space-time correlations

$$J_{ij}(\mathbf{r}_1,\mathbf{r}_2) = \langle E_i^*(\mathbf{r}_1,t)E_j(\mathbf{r}_2,t)\rangle, \; i = x, y; j = x, y \quad (3)$$

Here $E_x$ and $E_y$ are the components of the complex electric field vectors in x and y directions, respectively and $t$ is the time variable. The field in space-time domain is also assumed to be wide-sense stationary and ergodic, and the angular brackets denote the time average. Clearly, both these representations yield that the elements of these matrices are correlation function and are in general quantities of complex nature.

    Both the CSD and BCP matrices possess four elements each, characterized by the correlations between components of the electric fields with two spatial arguments. Although the space-frequency treatment provides another direction for determining the change in various frequency dependent parameters of the random electromagnetic beams, yet the space-time based treatment remains more fundamental in nature [43, 44]. It provides comparatively easier and adequate means for the analysis of the fundamental phenomena such as interference and diffraction of light and, also in situations where time dependent properties of the optical fields are of interest [28, 29]. However, in case the light field is monochromatic or sufficiently narrow band, both the characterizations can be treated similar [45]. In the framework of experiments in which a quasi-monochromatic beam (such as alaser beam) is used, both the CSD and BCP matrices yield identical results [46].

## 3. Two-point Stokes parameters of EM fields
Similar to usual Stokes parameters, which can be represented as linear combinations of the elements of coherency matrix, the linear combination of elements of two-point correlation matrices (CSD or BCP) yield the two-point or generalized Stokes parameters [47]. In space-frequency domain, these parameters can be expressed in terms of the elements of the CSD matrix as

$$S_0(\mathbf{r}_1,\mathbf{r}_2,\omega) = \langle E_x^*(\mathbf{r}_1,\omega)E_x(\mathbf{r}_2,\omega)\rangle + \langle E_y^*(\mathbf{r}_1,\omega)E_y(\mathbf{r}_2,\omega)\rangle, \qquad 4(a)$$

$$S_1(\mathbf{r}_1,\mathbf{r}_2,\omega) = \langle E_x^*(\mathbf{r}_1,\omega)E_x(\mathbf{r}_2,\omega)\rangle - \langle E_y^*(\mathbf{r}_1,\omega)E_y(\mathbf{r}_2,\omega)\rangle, \quad\quad 4(b)$$

$$S_2(\mathbf{r}_1,\mathbf{r}_2,\omega) = \langle E_x^*(\mathbf{r}_1,\omega)E_y(\mathbf{r}_2,\omega)\rangle + \langle E_y^*(\mathbf{r}_1,\omega)E_x(\mathbf{r}_2,\omega)\rangle, \quad\quad 4(c)$$

$$S_3(\mathbf{r}_1,\mathbf{r}_2,\omega) = i\left[\langle E_y^*(\mathbf{r}_1,\omega)E_x(\mathbf{r}_2,\omega)\rangle - \langle E_x^*(\mathbf{r}_1,\omega)E_y(\mathbf{r}_2,\omega)\rangle\right]. \quad\quad 4(d)$$

In order to interpret the physical meaning of these parameters, one can take analogy from the standard interpretation of the usual Stokes parameters [7]. Following [23, 36], one can represent these parameters analogues to their single-point counterparts as

$$S_0(\mathbf{r}_1,\mathbf{r}_2,\omega) = W_{xx}(\mathbf{r}_1,\mathbf{r}_2,\omega) + W_{yy}(\mathbf{r}_1,\mathbf{r}_2,\omega), \quad\quad 5(a)$$

$$S_1(\mathbf{r}_1,\mathbf{r}_2,\omega) = W_{xx}(\mathbf{r}_1,\mathbf{r}_2,\omega) - W_{yy}(\mathbf{r}_1,\mathbf{r}_2,\omega), \quad\quad 5(b)$$

$$S_2(\mathbf{r}_1,\mathbf{r}_2,\omega) = W_{\alpha\alpha}(\mathbf{r}_1,\mathbf{r}_2,\omega) - W_{\beta\beta}(\mathbf{r}_1,\mathbf{r}_2,\omega), \quad\quad 5(c)$$

$$S_3(\mathbf{r}_1,\mathbf{r}_2,\omega) = W_{rr}(\mathbf{r}_1,\mathbf{r}_2,\omega) - W_{ll}(\mathbf{r}_1,\mathbf{r}_2,\omega), \quad\quad 5(d)$$

where $\alpha$ and $\beta$ represent the components in a coordinate system that are rotated 45° and 135° in a counter-clockwise direction, and $r$ and $l$ represent the components having the right and the left-hand circular polarizations, respectively. Since in Eq. (5), we are now dealing with correlations, the physical interpretation is the following: the parameter $S_0$ describes the sum of the electric-field $x$ and $y$ component correlations at points $\mathbf{r}_1$ and $\mathbf{r}_2$, whereas parameters $S_1$, $S_2$, and $S_3$ describe the differences of the correlations in the orthogonal electric-field $x$, $y$, and $\alpha$, $\beta$, and $r$, $l$ components, respectively. Clearly two-point (generalized) Stokes parameters also represent the spatial coherence properties of different polarization components of the light fields and being correlation functions, these parameters are in general complex quantities [15].

## 4. Electromagnetic spectral interference law

It is well known that in scalar optics, lack of intensity fringes at the observation plane is a direct manifestation of null correlation between interfering beams; i.e., the field at the two pinholes in Young's experiment is incoherent. Since for producing best contrast interference fringes, polarization of interfering fields need to be parallel, absence of intensity modulation does not always imply that the electric field at the two pinholes lacks coherence [27]. For example, in the electromagnetic case the interference manifests not always as intensity fringes but sometimes in the form of polarization modulations also. Thus in order to conclude the coherence properties of EM beams, one needs to place coherence and polarization at the same footings [30].

The connection between usual (single-point) Stokes parameters and the generalized (two-point) Stokes parameters can be understood in an intuitive manner by using the recently proposed Electromagnetic Spectral Interference Law [27, 48]. The usual Stokes parameters $S_n(\mathbf{r}, \omega)$, for $n = 0 \ldots 3$ due to the two combined beams at point P in the plane O (refer to Fig. 1) can be expressed in terms of the Stokes parameters due to the individual beams $S_n^{(i)}(\mathbf{r}, \omega)$, $i = 1, 2$ as

$$S_n(\mathbf{r},\omega) = S_n^{(1)}(\mathbf{r},\omega) + S_n^{(2)}(\mathbf{r},\omega) + 2 \cdot \left[S_0^{(1)}(\mathbf{r},\omega)\right]^{1/2} \cdot \left[S_0^{(2)}(\mathbf{r},\omega)\right]^{1/2} \cdot |\mu_n(\mathbf{r}_1,\mathbf{r}_2,\omega)|$$
$$\times \cos\left\{\arg[\mu_n(\mathbf{r}_1,\mathbf{r}_2,\omega)] - \frac{2\pi}{\lambda}(R_1 - R_2)\right\}, \quad\quad (6)$$

where $|\mu_n(\mathbf{r}_1,\mathbf{r}_2,\omega)|$ is the absolute value and $\arg[\mu_n(\mathbf{r}_1, \mathbf{r}_2, \omega)]$ is the phase of the complex quantity $\mu_n(\mathbf{r}_1,\mathbf{r}_2,\omega)$, respectively, and the last term of cosine function is due to the path difference in distances between the pinholes/slits and the observation point. Eq. (6) shows that, in general, the usual Stokes parameters of light at a point P in the observation plane are the sum of the Stokes parameters due to the two beams reaching P from the two slits and an additional term on the right hand side. In case the observation point is an axial point (to the best extent possible), for all practical purposes we can treat $R_1 \approx R_2$. Due to this, the path dependent term in Eq. (6) vanishes. The additional term consists of the modulation parameters $\mu_n(\mathbf{r}_1,\mathbf{r}_2,\omega)$, which can be expressed in terms of the normalized generalized Stokes parameters $S_n(\mathbf{r}_1,\mathbf{r}_2,\omega)$, as

$$\mu_n(\mathbf{r}_1, \mathbf{r}_2, \omega) = \frac{S_n(\mathbf{r}_1,\mathbf{r}_2,\omega)}{[\varphi(\mathbf{r}_1,\omega).\varphi(\mathbf{r}_2,\omega)]^{1/2}}, \quad\quad (7)$$

where $\varphi(\mathbf{r}_i, \omega) = \varphi_x(\mathbf{r}_i, \omega) + \varphi_y(\mathbf{r}_i, \omega)$ for (i=1, 2) are the spectral densities or equivalently the first Stokes parameter $S_0(\mathbf{r}_i, \omega)$ at slits $\rho_1$ and $\rho_2$, respectively (see Fig. 1). Since $S_n(\mathbf{r}_1,\mathbf{r}_2,\omega)$ can be expressed in terms of the elements of CSD matrix, we can see that the linear combination of terms $\eta_{ij}(\mathbf{r}_1, \mathbf{r}_2, \omega)$ corresponding to $\mu_n(\mathbf{r}_1, \mathbf{r}_2, \omega)$ are the normalized CSD matrix elements [Eq. (1)], which characterize the field correlations at the pinholes [27], as

$$\eta_{ij}(\mathbf{r}_1, \mathbf{r}_2, \omega) = \frac{W_{ij}(\mathbf{r}_1,\mathbf{r}_2,\omega)}{[\varphi(\mathbf{r}_1,\omega).\varphi(\mathbf{r}_2,\omega)]^{1/2}}. \tag{8}$$

In Eq. (8), $\eta_{ij}(\mathbf{r}_1, \mathbf{r}_2, \omega)$ are a measure (quantification) of correlations between all sets of electric field components. For scalar fields, this quantity is referred as spectral degree of coherence, whereas for vector fields, we have 4 such correlations, which are essentially to be known to predict the degree of coherence for the EM fields. For equal intensities at the pinholes, i.e. $S_0^{(1)}(\mathbf{r}, \omega) = S_0^{(2)}(\mathbf{r}, \omega)$, at the double-slit plane, one can define the modulus of normalized two-point Stokes parameters as a contrast parameter as

$$V_n(\omega) = |\mu_n(\mathbf{r}_1, \mathbf{r}_2, \omega)| \text{ for n=0..3}, \tag{9}$$

where $0 \leq V_n(\omega) \leq 1$. Thus, one can see that coherence between any set of the field components at the pinholes results in the modulation of at least some Stokes parameter.

## 5. Electromagnetic degree of coherence

Since the EM field can have any degree of coherence and polarization, correlations corresponding to all four sets of the components of electric fields needs to be determined [30, 49]. These correlations can be quantified in terms of respective degrees of coherence and a combination of all the four degrees of coherence give the electromagnetic degree of coherence in space-frequency domain as [22, 27]

$$\mu^2(\mathbf{r}_1,\mathbf{r}_2,\omega) = \frac{\mathrm{tr}[W(\mathbf{r}_1,\mathbf{r}_2,\omega).W(\mathbf{r}_2,\mathbf{r}_1,\omega)]}{\langle S(\mathbf{r}_1,\omega)\rangle\langle S(\mathbf{r}_2,\omega)\rangle}, \tag{10}$$

where $S(\mathbf{r}_i,\omega) = W(\mathbf{r}_i,\mathbf{r}_i,\omega)$ is the spectral density at a point $\mathbf{r}_i$. Definition of EM degree of coherence given by Eq. (10) account for the modulation in all four Stokes parameters. Using Eq. (9), one can show that EM degree of coherence can be expressed in terms of contrast parameters as [30]

$$\mu^2(\mathbf{r}_1,\mathbf{r}_2,\omega) = \frac{1}{2}\sum_{n=0}^{3} V_n^2(\omega). \tag{11}$$

Eq. (11) provides the physical interpretation of EM degree of coherence being a direct measure of the contrasts of modulation in the one-point Stokes parameters which is analogous to that of the scalar degree of coherence, which describes only the visibility of the intensity fringes on the observation screen.

In analogy with abovementioned coherence functions, which represent beam properties at two-space points, one can have correlations functions characterizing two-time points [38]. Such temporal correlation properties of EM beams have also been a center of research during past decade. The temporal EM degree of coherence was investigated using the EM temporal interference law in a temporal coherence-based interferometer (Michelson interferometer), and two-point temporal coherence properties were proposed and determined experimentally [39, 40]. Very recently the EM longitudinal spatial coherence for laser beam has been investigated both theoretically and experimentally showing a drastic change in longitudinal coherence length by variation in angular spectrum of the source which is applicable in several domains of optical research [41, 50, 51].

Model sources such as gaussian Schell-model (GSM) exhibit rich coherence features [12]. They generate beams which are having well defined characteristics and resemble to natural light sources. Very recently electromagnetic version of these sources namely electromagnetic gaussian Schell-model (EGSM) sources received increased attention [52-57]. Beams generated by such sources (EGSM beams) have been studied and their generation and characterization methods have been proposed and demonstrated experimentally [31, 58, 59].

## 6. Measurement of two-point spatial coherence matrix

The two-point spatial coherence functions have been measured in the last decade experimentally both in the space-time and in space-frequency domain. These methods include use of first order interference and second order interference for the determination of coherence-polarization features of the EM beams. The first theoretical proposal for the measurement of elements of 2x2 electric cross-spectral density has been made by Roychowdhury and Wolf in 2003, in which a procedure was described using Young's interferometer with the help of two polarizers and a rotator [60]. Using different polarizing elements before the separate slits, the components of the cross-spectral density matrix $\vec{W}(\mathbf{r}_1,\mathbf{r}_2,\omega)$ were given by

$$\vec{W}(\mathbf{r}_1,\mathbf{r}_2,\omega) = W_{ij}(\mathbf{r}_1,\mathbf{r}_2,\omega) = \sqrt{S_i(\mathbf{r}_1,\omega)}\sqrt{S_j(\mathbf{r}_2,\omega)}\eta_{ij}(\mathbf{r}_1,\mathbf{r}_2,\omega), \quad (12)$$

where $(i,j = x,y)$. This proposal was realized experimentally for an expended laser beam using a modified version of the Young's interferometer [32]. As shown in Fig. 2, laser beam from a randomly polarized Helium-Neon laser (make Melles Griot) was expanded using a spatial filter beam expander assembly and was passed through a symmetrical double-slit of slit width 150μm and slit separation 200μm, placed in the beam path at 30cm away from the expander as shown in Fig. 2(a). Two right angle front coated prisms were introduced to separate the two beams by approximately 8cm and recombined them again at some distance away from the first prism. The interference fringes were obtained at the observation plane at a distance 135cm from the second prism and were photographed as shown in Fig. 2. A fibre coupled spectrometer mounted with a computer controlled motorized micro-positioner was used to measure the spectra at the observation plane.

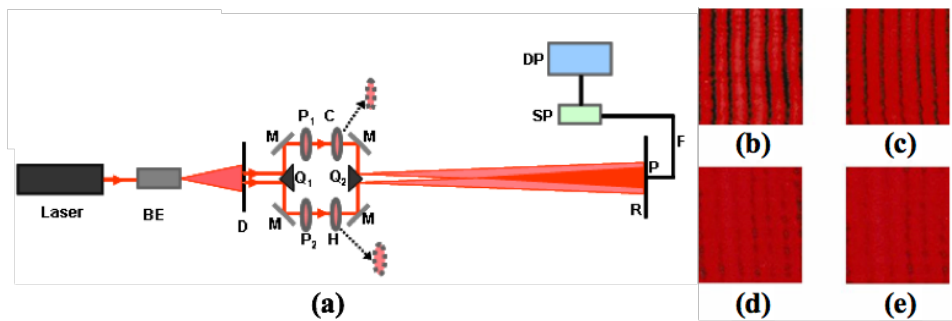

Fig 2. (a) Experimental setup for the measurement of two-point correlations of laser field using a modified Young's scheme. Notations: BE beam expender, Q prism, M mirror, P polarizer, H half wave plate, C compensator plate, R observation plane, SP spectrometer F optical fibre and DP data processor, (b)-(e) Interference fringes obtained for different combinations of polarizer and half wave plate [32].

To calculate $W_{xx}(\mathbf{r}_1,\mathbf{r}_2,\omega)$ [$W_{yy}(\mathbf{r}_1,\mathbf{r}_2,\omega)$] component of the electric cross-spectral density matrix, two identical dichroic sheet polarizers having direction of polarization along the x-axis [y-axis] were inserted in the separate beam paths (Fig. 1). This allowed the x [y] components of the beams only to pass through and good visibility interference fringes were obtained at R as shown in Fig. 2(b) [2(c)]. The maximum and minimum values of the spectral density were recorded by the spectrometer tracing the fibre tip horizontally over the central interference fringe. The spectral densities at a point P over the central fringe due to the individual slits (points $Q_1$ and $Q_2$) were obtained by making measurements when only one slit was open and other closed. $W_{xy}(\mathbf{r}_1,\mathbf{r}_2,\omega)$ [$W_{yx}(\mathbf{r}_1,\mathbf{r}_2,\omega)$] component of the matrix was obtained when polarization of $P_1$ polarizer was in x-axis while for $P_2$ was in y-axis. To bring the vibrations in the same plane, so that interference could be possible, a half wave plate (H) having optics axis at $45^0$ with the incident polarization of light was placed after the polarizer $P_2$ which worked as a $90^0$ polarization rotator. Another half wave plate C with identical specifications having optic axis along the incident polarization of light was placed in another arm for path compensation [Fig. 2(a)]. The interference fringes thus obtained are shown in Fig. 2(d) [2(e)]. With one slit closed and the other open, the spectral densities due to the individual slits were also recorded at the same point P. It is evident from the photographs of Fig. 2 that the fringe visibility due to xy and yx components is much less than the xx and yy components. The maximum and minimum values of the spectral density for the same pair of points were recorded by the spectrometer at the observation plane for the central fringe and the spectral degree of coherence was calculated using the formula [18]

$$\eta(\mathbf{r}_1,\mathbf{r}_2,\omega) = \frac{S_{max} - S_{min}}{S_{max} + S_{min}}, \quad (13)$$

where $S_{max}$ and $S_{min}$ were the values of spectral density around the central fringe when both the slits were open. These visibilities were measured for all the four sets of components and using Eq. (12), the elements of the CSD matrix at an axial point were thus obtained as

$$\vec{W}(\mathbf{r}_1,\mathbf{r}_2,\omega) = \begin{bmatrix} 410.6 \pm 4.0 & 118.2 \pm 1.6 \\ 119.7 \pm 1.8 & 366.4 \pm 3.4 \end{bmatrix}. \quad (14)$$

In the abovementioned experiment [32], the spectral measurements were taken at an axial point in the observation plane, where the fringe intensity was having a maxima. This reduced the complex coherence parameters into real ones. The method is quite useful in the laboratory scale, however, practical situations could be otherwise, i.e. either the observation point could be an off-axis point or the light could be partially coherent. These conditions turn the CSD matrix elements as coherence functions, which are complex quantities [35]. Thus one needs to measure both the amplitude (modulus) and phase of these elements experimentally.

In order to determine the phase part of the spectral degree of coherence, the spectral interference law proposed in [27] was used. For the condition when spectral densities at point P due to both the beams were approximately identical, one obtains

$$\mathrm{Re}\,\eta(\mathbf{r}_1,\mathbf{r}_2,\omega) = \frac{\varphi(\mathbf{r},\omega) - 2\varphi^1(\mathbf{r},\omega)}{2\varphi^1(\mathbf{r},\omega)}, \quad (15)$$

where the real and imaginary parts of the spectral degree of coherence can be expressed as

$$\mathrm{Re}\,\eta(\mathbf{r}_1,\mathbf{r}_2,\omega) = |\eta(\mathbf{r}_1,\mathbf{r}_2,\omega)|\cos\beta(\mathbf{r}_1,\mathbf{r}_2,\omega),$$
$$\mathrm{Im}\,\eta(\mathbf{r}_1,\mathbf{r}_2,\omega) = |\eta(\mathbf{r}_1,\mathbf{r}_2,\omega)|\sin\beta(\mathbf{r}_1,\mathbf{r}_2,\omega). \quad (16)$$

Using Eqs. (15) and (16), the complex value of the spectral degree of coherence is thus obtained as [32]

$$\eta(\mathbf{r}_1,\mathbf{r}_2,\omega) = \nu(\mathbf{r}) \times \left[ \left\{ \frac{\mathrm{Re}\,\eta(\mathbf{r}_1,\mathbf{r}_2,\omega)}{\nu(\mathbf{r})} \right\} + i \left\{ \frac{\sqrt{\{\nu(\mathbf{r})\}^2 - \{\mathrm{Re}\,\eta(\mathbf{r}_1,\mathbf{r}_2,\omega)\}^2}}{\nu(\mathbf{r})} \right\} \right]. \quad (17)$$

The quantities inside the square brackets in Eq. (17) are real (cosine) and imaginary (sine) values of the spectral degree of coherence, which were obtained using experimental data for all the four sets of components. The quantities $\nu_{ij}(\mathbf{r})$ and $\eta_{ij}(\mathbf{r}_1,\mathbf{r}_2,\omega)$ for $(i,j)=(x,y)$, were determined experimentally by placing polarizers and half wave plates in front of the individual slits with appropriate orientation and by taking spectral measurements at and around the off-axis point in the observation plane. These values were put in Eq. (12) giving the complex values of the elements of electric cross-spectral density matrix for a pair of points in the cross-section of the electromagnetic beam, shown in **Table 1** below.

Table 1. Complex elements of the electric cross-spectral density matrix [32]

| | |
|---|---|
| $W_{xx}(\mathbf{r}_1,\mathbf{r}_2,\omega)$ | (622±9.8)+i(353.7±6) |
| $W_{xy}(\mathbf{r}_1,\mathbf{r}_2,\omega)$ | (-115.3±4)+i(210±6.4) |
| $W_{yx}(\mathbf{r}_1,\mathbf{r}_2,\omega)$ | (-193±5.3)+i(108±4.4) |
| $W_{yy}(\mathbf{r}_1,\mathbf{r}_2,\omega)$ | (485±8.1)+i(369±6.5) |

Similar to the measurement of space-frequency two-point spatial correlations, one can have two-point spatial correlations in space-time domain also [16]. In space-time domain, an experimental method for determining the complex four elements of BCP matrix, has been demonstrated in [33]. The magnitude and phase of the complex degree of cross-correlation were determined experimentally by inserting polarizers and half-wave retarders in the two separate paths of a modified version of the Young's interferometer. The scheme was essentially similar to shown in Fig. 2 (a), however, instead of using spectral measurement device, a photodetector was used for time domain measurements. These experimentally determined values of the complex degree of cross-correlation, as shown in Fig. 3, were used with the mathematical expressions presented in the paper to determine the elements of the BCP matrix. Laser beam was taken as an electromagnetic field due to its wide optical applications.

Spatial coherence functions have been measured in past for a variety of experimental settings. This include use of non-parallel double-slits [61], reversed-wavefront Young's interferometer [62], using digital micromirror [63], using diffraction [64] etc. Coherence functions of several kinds of beam shapes and distributions have been investigated and measured, which mainly include complex valued objects [65], expended laser beams [32], EMGSM beams [31], partially coherent broadband beam [36], vortex beam [66], radially polarized EM beam [67], focused vortex beams [68, 69] and very recently for vector vortex beams [70]. Very recently coherence function was also measured using nano-scattering [42].

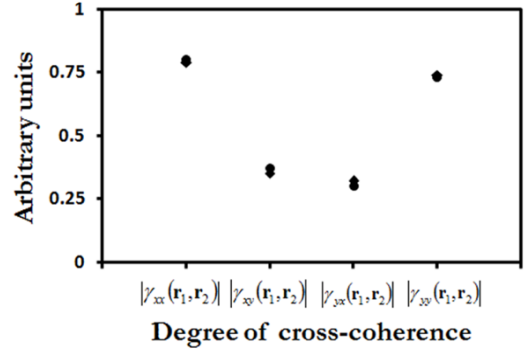

Fig 3: A comparison between the theoretical expected and experimentally obtained absolute values of the degree of cross-coherence showing an excellent agreement. Triangles represent the values measured experimentally while full circles show the corresponding theoretically expected values as detailed in [33].

7. **Measurement of two-point Stokes parameters**

As explained earlier, similar to 2x2 CSD and BCP matrices, two-point (generalized) Stokes parameters in space-frequency and in space-time domain also provide the coherence properties of different polarization components of light fields. These parameters can be expressed as linear combinations of elements of CSD (BCP) matrix, in space-frequency (space-time) domain, as shown in Eq (5). Based on this dependence, two-point Stokes parameters can be determined directly from the experimentally measured values of CSD matrix elements. In above-mentioned studies, in which these two-point correlation metrices were determined, the corresponding two-point Stokes parameters were also obtained. For example, in [34], the two-point Stokes parameters determined using Eqs. (6)-(11), are expressed in **Table 2**. One can clearly see that these parameters are complex functions and provide the information of spatial coherence between different polarization components of light fields.

Table 2: Two-point Stokes parameters obtained from the CSD matrix [34]

| | |
|---|---|
| $S_0(\mathbf{r}_1,\mathbf{r}_2,\omega)$ | $(1107\pm13)+i(722.7\pm8.3)$ |
| $S_1(\mathbf{r}_1,\mathbf{r}_2,\omega)$ | $(137\pm13)+i(-15.3\pm8.3)$ |
| $S_2(\mathbf{r}_1,\mathbf{r}_2,\omega)$ | $(308.3\pm6.4)+i(318\pm7.1)$ |
| $S_3(\mathbf{r}_1,\mathbf{r}_2,\omega)$ | $(102\pm7.1)+i(-78\pm6.4)$ |

Since determination of CSD matrix elements and two-point Stokes parameters using the modified version of Young's interferometer was a complex procedure due to precise alignment and narrow separation of two-beams after the double-slit, another method to determine directly the generalized Stokes parameters for a pair of points in the cross-section of an expanded laser beam making use of the electromagnetic spectral interference law was proposed and experimentally implemented [22, 34]. In this method, the generalized Stokes parameters for a pair of points in the cross-section of an expanded laser beam were determined experimentally using the usual Stokes parameters and visibility measurements. The experimental setup consisted of a randomly polarized He-Ne laser (make Melles Griot). The laser beam was expanded and was made incident on a double-slit having slit width 150 μm and slit separation 200 μm. The two beams emerging out from the double-slit interfered and interference fringes were obtained in plane R, at a distance 150 cm from the slit, where linear fringes were obtained. The spectral measurements were taken around an off-axis point P(**r**) using a fibre coupled spectrometer interfaced with a personal computer. The fringe pattern across the observation plane was scanned mounting the fibre tip on a micro-positioner.

The single-point Stokes parameters were determined experimentally by using a pair of optical elements in the conventional Young's interferometer by taking spectral measurements at point P and using the following expressions

$$S_0^1(\mathbf{r},\omega) = \varphi(0^0,0^0) + \varphi(90^0,0^0), \tag{18a}$$

$$S_1^1(\mathbf{r},\omega) = \varphi(0^0,0^0) - \varphi(90^0,0^0), \tag{18b}$$

$$S_2^1(\mathbf{r},\omega) = \varphi(45^0,0^0) - \varphi(135^0,0^0), \tag{18c}$$

$$S_3^1(\mathbf{r},\omega) = \varphi(45^0,45^0) - \varphi(135^0,45^0), \tag{18d}$$

where $\varphi(\theta,\phi)$ is the spectral density at P when the polarizer makes angle $\theta$ with the x axis and $\phi$ is the orientation of optic axis of the quarter wave plate with the incident polarization of light (the axis of the polarizer). Taking a general case in which the observation point was an off-axis point and spectral measurements were taken in and around that point, the contrast parameters and subsequently the two-point Stokes parameters were determined for the expended laser beam, which are shown in **Table 3**.

Table 3. The contrast parameters $C_n(r,\omega)$ and the generalized Stokes parameters $S_n(r_1,r_1,\omega)$ determined for an expanded laser beam [34]

| | | | |
|---|---|---|---|
| $C_0(\mathbf{r},\omega)$ | 0.79 ± 0.08 | $S_0(\mathbf{r}_1,\mathbf{r}_2,\omega)$ | (600 ± 19.2) + (930 ± 26.0) i |
| $C_1(\mathbf{r},\omega)$ | 0.14 ± 0.02 | $S_1(\mathbf{r}_1,\mathbf{r}_2,\omega)$ | (11 ± 0.55) + (195 ± 10.6) i |
| $C_2(\mathbf{r},\omega)$ | 0.15 ± 0.04 | $S_2(\mathbf{r}_1,\mathbf{r}_2,\omega)$ | (19 ± 0.87) + (207 ± 12.1) i |
| $C_3(\mathbf{r},\omega)$ | 0.20 ± 0.07 | $S_3(\mathbf{r}_1,\mathbf{r}_2,\omega)$ | (257 ± 12.7) + (97 ± 9.6) i |

These above-mentioned methods are two-step processes in which one needs to determine either the CSD matrix or the one-point Stokes parameters and subsequently the two-point Stokes parameters are determined. In one of the more recent studies [23] it has been proposed that the generalized Stokes parameters have physical interpretation as linear combinations of cross-spectral density functions of specific electric field components, as shown in Eq. (5) [10]. This is fully analogous to the classical intensity based interpretation of the conventional Stokes parameters. This analogy in these two kinds of Stokes parameters advocates for an experimental approach for determining the two-point Stokes parameters similar to the one used for determining the single-point (conventional) Stokes parameters. Additionally so far, these two-point coherence functions were determined only for a laser source, having stimulated emission properties, high monochromaticity and coherence. Natural light beams having spontaneous emission features are partially coherent and quasi-monochromatic, and two-point coherence properties of such beams are needed to be probed.

To realize the proposed method experimentally [36], the schematic diagram is shown in Fig. 4. A continuous spectrum tungsten-halogen lamp L (Mazda, colour temperature = 3200K) operated at 600W using a regulated dc power supply (Heinzinger, stability 1 part of 104) was used with diffused outer glass jacket of the lamp to obtain uniform illumination. A glass filter F having peak wavelength at 580nm and spectral width of 90nm was used to filter the polychromatic beam giving out broadband optical field. To obtain higher visibility of the interference fringes, the partial coherence of the light beam was enhanced by first passing the beam through a narrow single slit (SS) having width 400μm. The radiation from SS was then made incident on a double slit DS (slit width=250μm, slit separation=450μm) placed at a distance 100cm from the single-slit, symmetric to the optical axis (see Fig. 1). The double-slit is used here to probe the beam at two different points in its cross-section. The active area of the double slit (700μm) was well occupied within the partial coherence region developed at DS (using the van-Cittert Zernike theorem with SS), which was calculated to be 1.4 mm for present geometry.

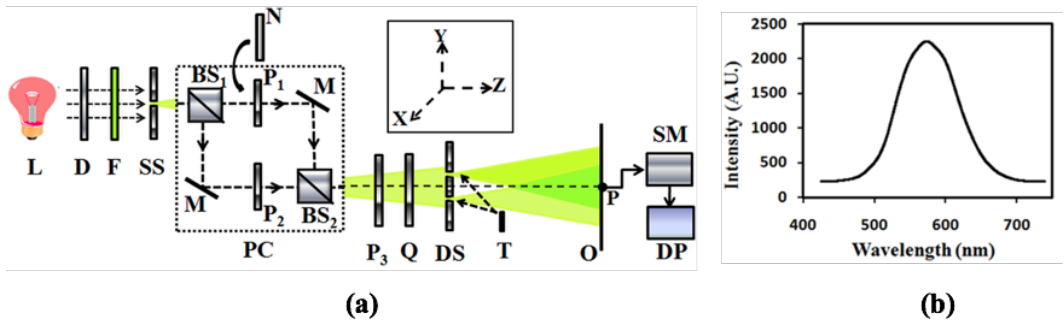

Fig 4. (a) Schematic experimental setup consisting of a variable degree of polarization source. Notations L tungsten halogen lamp, D diffuser, F broadband filter, SS single-slit, PC polarization controller consisting of BS beam-splitter, M mirror, P polarizer and N neutral density filter; Q quarter wave plate, DS double-slit, T stopper, O observation plane, SM spectrometer and DP data processor. (b) Spectral profile of filtered white light source showing broadband (90nm bandwidth) nature [36].

The state of polarization of light beam was controlled before DS by passing the beam through a polarization controller PC, as shown in Fig. 4(a). Inside the polarization controller, to facilitate the insertion of polarizers (P1, P2) and neutral density filter N in the beam paths, the beam was first divided into two beams of nearly equal intensity using a 50:50 beam-splitter BS1. After reflecting through the mirrors M and travelling equal paths, these two beams were recombined using another beam-splitter BS2, which actually worked as beam combiner. A set of polarizer P3 and quarter wave plate Q was used to choose proper basis of the coordinate systems. A fiber-coupled spectrometer SM connected with a personal computer DP was used for spectral measurements. The spectral density of the beam passing through any of the slits is shown in Fig. 4(b).

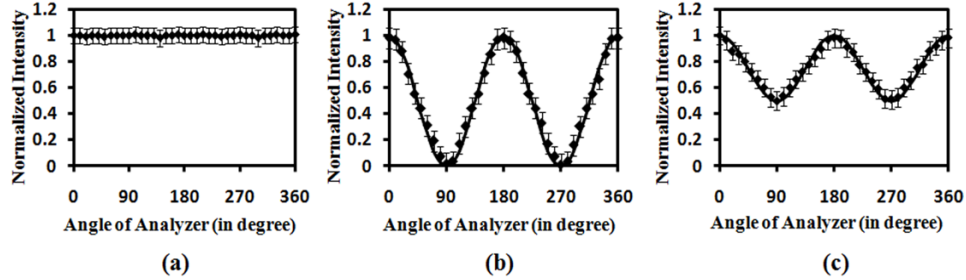

Fig 5. Characterization of different degrees of polarizations of the secondary source showing (a) unpolarized light, (b) linearly polarized light in x direction and (c) partially polarized light. The error bars show uncertainty in the measurement [36].

The unpolarized light, which can be expressed as a combination of two equal-intensity orthogonally polarized light beams having no mutual correlation, was realized using polarizers $P_1$ and $P_2$ in different beams (in PC) making angles $0^0$ and $90^0$ with x-axis respectively. The recombined beam was passed through another polarizer ($P_3$) and the output intensity after showed a flat response at all polarizer angles (Fig. 5a) showing the unpolarized nature of light beam. To produce linearly polarized light beam, both $P_1$ and $P_2$ were put with their axis in x-direction. For this beam, the output intensity followed the Malus law [2], as shown in Fig. 5(b) showing the linearly polarized nature of the light beam. To form partially polarized state of light beam, $P_1$ was replaced by a neutral density filter N (transmittance 50%) in PC of Fig. 4. Not only does this allow the unpolarized light to pass through but also puts its transmitted intensity at par with the transmitted intensity of $P_2$. As a result, the recombined beam consists of linearly polarized and unpolarized light with equal proportion of intensity, giving a state of partial polarization. This partially polarized nature of light beam was confirmed when the output intensity was plotted as a function of the angle of analyser. As illustrated in Fig. 5(c), the light intensity fluctuates and does not have a zero as minima. For all these three types of polarizations, spectral measurements were made at the observation plane for different combinations of $P_3$ and Q, and the four two-point Stokes parameters were determined, which are shown in **Table 4**.

**Table 4 Experimentally determined values of the four two-point Stokes parameters for different polarization states of the light beam[36]**

|  | **Unpolarized light** | **Linearly polarized light** |
|---|---|---|
| $S_0(\mathbf{r}_1,\mathbf{r}_2,\omega)$ | (134.24±0.28)+(703.55±0.58)i | (136.48±0.25)+(561.10±0.33)i |
| $S_1(\mathbf{r}_1,\mathbf{r}_2,\omega)$ | (1.64±0.28)+(4.81±0.28)i | (135.56±0.25)+(449.72±0.23)i |
| $S_2(\mathbf{r}_1,\mathbf{r}_2,\omega)$ | (-0.41±0.01)-(12.68±0.11)i | (-1.46±0.02)-(2.99±0.03)i |
| $S_3(\mathbf{r}_1,\mathbf{r}_2,\omega)$ | (-1.96±0.02)-(57.61±0.17)i | (2.46±0.02)+(7.38±0.04)i |

|  | **Partially polarized light** |
|---|---|
| $S_0(\mathbf{r}_1,\mathbf{r}_2,\omega)$ | (137.32±0.26)+(622.06±0.41)i |
| $S_1(\mathbf{r}_1,\mathbf{r}_2,\omega)$ | (66.59±0.26)+(317.42±0.31)i |
| $S_2(\mathbf{r}_1,\mathbf{r}_2,\omega)$ | (-0.63±0.02)-(31.99±0.12)i |
| $S_3(\mathbf{r}_1,\mathbf{r}_2,\omega)$ | (-2.51±0.03)+(0.81±0.01)i |

From **Table 4**, one can see that in the first case, the last three generalized Stokes parameters $S_n(\mathbf{r}_1,\mathbf{r}_2,\omega)$ for $n=1,2,3,$ show a little correlation between different polarization components. This signifies the unpolarized nature of the light beam. For the second case, on the other hand, the total correlation given by first generalized Stokes parameter is the sole contribution of the linear polarization of light at the double-slit. The positive value of this quantity determines the dominance of x-linear polarization over the y-linear polarization. For the third case, the major contribution in total correlation is furnished by the x-axis linear polarization, as obvious by the positive value of the second two-point Stokes parameter. However, the remaining contribution is offered by the unpolarized part of the light beam. This infers that the probed beam is having partially polarized nature. The unique feature of this experimental method is minimum uncertainty in the measurements due to minimal optical components and minimal time taken.

8. **Measurement of two-point temporal coherence function**

Similar to correlations between two-spatial points, correlations between two-time points also play a vital role in coherence-polarization studies. Temporal electromagnetic degree of coherence (TEMDOC) and temporal degree of cross-polarization (TDOCP) define the coherence properties of different polarization components of the field at two points in time [40]. The absolute nature of these quantities offers a way to determine them by measuring correlations in intensity fluctuations [37]. In comparison with amplitude interferometers, the intensity interferometer is more robust to misalignment and phase fluctuations as it does not require interferometric phase stabilization [71]. Thus determination of these two-time correlations can be offered by intensity correlations in an easier manner. The classic Hanbury Brown and Twiss (HBT) interferometer provides a method to investigate correlations between intensity fluctuations in random EM fields at different points of space, and/or time [72]. For optical fields, use of intensity interferometer may provide a direct access to the coherence-polarization features [71] and the benefit of this interferometer is its insensitivity to phase errors. In the recent past, it has been shown that the random EM fields do possess intensity fluctuations, which properties were studied in great detail in connection with partially coherent and partially polarized fields [73-76].

The experimental scheme first required realization of a tunable DOP source [40]. Out of several such realizations in the recent past [36, 49, 77, 78], we aim to use an unbalanced Mach Zehnder interferometer (UMZI), in which the difference of arms is three times larger than the coherence length of the sources and it is used to construct a tunable DOP source, as shown in Fig. 6(a) [41]. Two kinds of light sources are used: 1) an unpolarized highly multimode spatially coherent, monochromatic laser ($\lambda$ = 650 nm, coherence length ≈ 25 cm), and 2) a broadband, partially coherent light source which is constructed using a white-light LED, apertures (A) and a lens (L) to select a collimated beam with narrow cross-section and filter the spectrum by using a bandpass filter (F, $\lambda 0$ = 550 nm, $\Delta\lambda$ = 70 ± 10 nm). Fig. 6(b) shows the drastic variation in spectral properties of these sources. Beams from both the sources pass through the first polarizing beam splitter (PBS) which divides it in orthogonal polarizations. Use of a polarizer (P) having $45^0$ angle with $x$ axis placed before the PBS assures equal intensities of the two orthogonal polarizations. One of the two beams passes through a half wave plate (HWP) before mixing on a non-polarizing beam splitter (NPBS) by rotating which the degree of polarization of the source can be controlled [49]. The resultant field of known degree of polarization first passes through a set of Quarter wave plate (QWP) and polarizer (P) in order to select different field components to determine the two-time Stokes parameters. Finally the field enters the intensity interferometer consisting of a 50 : 50 NPBS and two detectors ($D1$, $D2$) [29]. One of the photodiodes can be translated axially in order to change the path difference of the interferometer. We emphasize that the alignment of this scheme is trivial compared to the amplitude based interferometer, which requires interferometric stabilization of phase for the measurement.

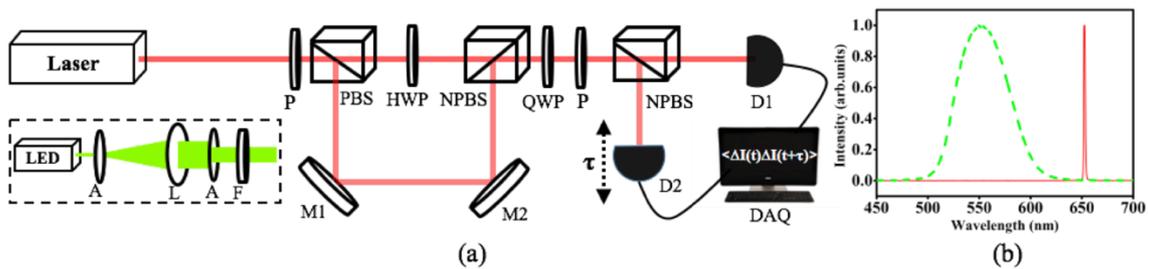

Fig 6. (a) Experimental scheme for the determination of TEDOC and TDOCP using intensity correlations. (Inset) LED based broadband partially coherent source. Abbreviations are described in the text. (b) Comparison of spectra of the laser (red line) and the filtered white-light LED (yellow–green dashed line) [40].

At the output of the UMZI, the DOP of light can be tuned as a function of the HWP placed in the interferometer. As shown in Fig. 7, the DOP variation with the angle of HWP is plotted for both the laser beam (a) and the filtered LED source (b). These variations are corrected for the polarization dependent transmission and reflection of the PBS used in the experiment, and the theoretical fit (in red) also includes these dependences. One can clearly see that in both the cases, the DOP can be continuously varied from near 0 to approximately 1, showing a source capable of arbitrarily producing near unpolarized, partially polarized and near polarized light beams, which may find potential applications in coherence-polarization studies.

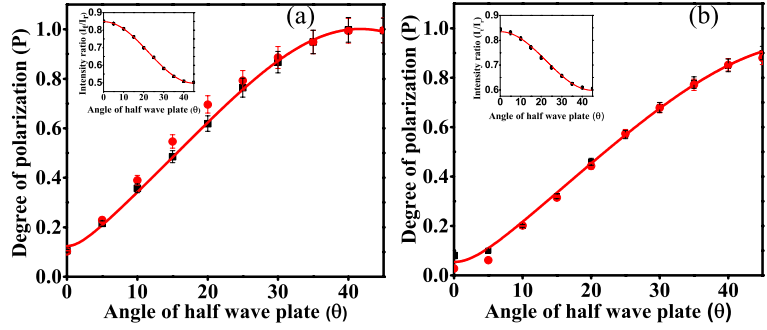

Fig. 7 (a) Variation in DOP of the laser beam by varying the HWP angle $\theta$ in Fig. 6. (b) Change in DOP of the LED beam with respect to the change in $\theta$. Black filled squares represent the experimental data obtained from Stokes measurements, red filled circles represent the experimental data obtained using intensity interferometer and the red line shows the theoretical fit. Inset shows the non-ideal behaviour of the NPBS of the Mach–Zehnder assembly [40].

For any partially polarized field propagating along z axis, the intensity correlations of this field is given by,
$$\Pi(\tau) = \sum_{i,j} |< E_i^*(t)E_j(t+\tau) >|^2 . \tag{19}$$
In the output of HBT interferometer, the intensity correlations in terms of the two-time Stokes parameters is given as,
$$\Pi(\tau, \theta) = \frac{1}{4} | S_0(\tau) + S_1(\tau)\cos^2 2\theta + S_2(\tau)\frac{\sin 4\theta}{2} + S_3(\tau)\sin 2\theta |^2 . \tag{20}$$

From Eq. (13) & (14) we got two-time Stokes parameters [40] as,

$$\begin{aligned}
S_0(\tau) &= \Pi(\theta_3)^{1/2} + \Pi(\theta_4)^{1/2}, \\
S_1(\tau) &= 2\Pi(\theta_1)^{1/2} - \Pi(\theta_3)^{1/2} - \Pi(\theta_4)^{1/2}, \\
S_2(\tau) &= 2\Pi(\theta_1)^{1/2} + (1+\sqrt{2})\Pi(\theta_3)^{1/2} + (1-\sqrt{2})\Pi(\theta_4)^{1/2} - 4\Pi(\theta_2)^{\frac{1}{2}}, \\
S_3(\tau) &= \Pi(\theta_3)^{1/2} - \Pi(\theta_4)^{\frac{1}{2}}.
\end{aligned} \tag{21}$$

Experimentally, we measured these two-time Stokes parameters. If $\tau=0$, these parameters one-time (usual) Stokes parameters. We calculated DOP, TDOCP and TEMDOC for the light beams. TEMDOC increases with degree of polarization of the source. But it decreases, when we increase the value of $\tau$, given by the following relation

$$\gamma^2(\tau) = \frac{1}{2} \frac{|S_0^2(\tau)|}{|S_0^2(0)|} [1 + P^2(\tau)] . \tag{22}$$

The results of Fig. 8 show that both the TEDOC and TDOCP increase with the DOP for all time separations within the coherence length and the extreme values of TDOCP are limited by the DOP of the corresponding sources [40]. A comparison of TEDOC values obtained for different path separations ($\tau \geq 0$) shows that for increasing the path difference, the TEDOC decreases, as the correlation between the time points decreases. TDOCP [Fig. 3(b)] remains nearly unperturbed with the change in the detector separation in the intensity interferometer (within coherence length) for both the sources and Its experimentally observed values are within the theoretically expected limits of 0 and 1 for any degree of polarization of the electromagnetic fields. The behaviour of TEDOC and TDOCP of the broadband LED field with respect to different DOPs for equal time separation of the detectors($\tau= 0$) is shown in Fig. 4. Since the coherence length of the broadband source was few microns, the resolution of the translation stage prevent us for making more measurements within coherence length for $\tau=0$ case. Fig. 4 shows that the equal time electromagnetic degree of coherence, i.e. TEDOC for $\tau= 0$ increases with DOP of the source, which is similar to the previous case. Analogously, the TDOCP at $\tau= 0$ for broadband LED linearly increases with DOP of the source. In fact in this situation TDOCP becomes DOP measured using intensity correlations, which naturally is identical to the one measured using field correlations, verifying our experimental observations. for non-stationary fields, which are having time dependent modulations, TDOCP is expected to depend more explicitly on time separations.

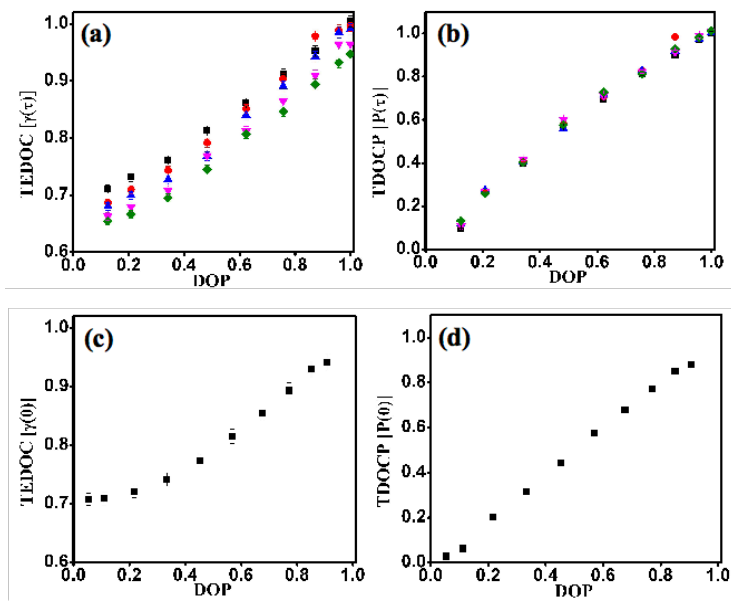

Fig 8. For the quasi-monochromatic laser source: (a) TEDOC as a function of the DOP, (b) TDOCP as a function of the DOP. In both (a) and (b) the difference in intensity interferometer arms, which results to time separation $\tau$, is given by: Black squares: 0 cm, red spheres: 3.0 cm, blue triangles: 6.0 cm, pink inverted triangles: 9.0 cm and green rotated squares: 12.0 cm. Plot of temporal parameters for the broadband LED source: (c) TEDOC as a function of the DOP, (d) TDOCP as a function of the DOP. In both (c) and (d) the measurements are made for the zero path difference of the interferometer arms ($\tau = 0$) [40].

9. **Applications of coherence functions**

As discussed earlier, two-point correlations for scalar optical fields essentially refer to spatial or temporal coherence properties. However, for EM fields which are of partially coherent and partially polarized nature, these correlations determine the partial coherence properties corresponding to different polarization components in spatial and temporal domains. One of the important results of partial coherence was coherence induced spectral changes studied mainly by E. Wolf and coworkers in $20^{th}$ century [79-82]. It has been shown both theoretically and experimentally that due to the fluctuating (or correlation) nature of light sources, a spectral shift of light can occur during its propagation even in free space called as "Wolf effect" [79]. Radiation from sources following the scaling law such as Lambertian sources do not show spectral change on propagation [80]. Determination of field correlations from spectral measurements led to a new technique called as spatial-coherence spectroscopy [82]. More recently, several attempts were made to explore hidden coherences and the connection between coherence, polarization and entanglement [83-87]. It was observed that visibility and distinguishability of light fields are connected with polarization of the light field [84]. These studies indicate towards the fuzziness of classical-quantum boundaries [76, 87].

High spatial and temporal coherence of laser beam produces speckle patterns which can be detrimental for many applications and appropriately chosen partially coherent fields can provide better performance in many of such cases [88]. These are several ways proposed to reduce the speckle patterns including use of slowly rotating ground glass diffuser, time-varying diffractive optical element, use of vibrating and multimode fibres etc [89-94]. Such low coherence beam has been used in holography and optical coherence tomography (OCT) [95, 96]. Imaging methods such as OCT require beams with lesser longitudinal (partial) spatial coherence, which increases axial resolution. OCT investigates change in the refractive index and the attenuation coefficient of samples such as living tissues, mostly to identify eye diseases or cancerous cells [97]. The figure of merit for OCT is the second-order cross-correlation function between the sample wave amplitude and the reference wave amplitude [98]. A method of tomography by adjusting the spatial coherence of light was also proposed and investigated [99].

One interesting approach which utilizes partially coherent light is speckle illumination for imaging (coherent) and microscopy [100-112]. Second-order correlations analysis, as already extensively used in imaging applications, moderately enhances the resolution. In biomedical optics, one of standard imaging methods is laser speckle contrast imaging, which is based on the 2nd-order correlation features [101-105]. Other methods such as fluorescence microscopy and super-resolution microscopy also have a basis on field correlations, which were also studies via methods such as intensity correlations and structured illumination [106-108]. Later on imaging methods such as structured illumination microscopy were developed based on partial coherence [109]. More recently image scanning microscopy is combined with quantum correlations resulting to super-resolution enhancement [110-112].

Ghost imaging (GI) is a technique that employs spatially correlated twin beams by correlating the outputs of two photodetectors to form the image of an unknown object [113-116]. The imaging resolution is provided by the beam that has never interacted with the object and by evaluating the spatial intensity correlation function between the two signals one can determine the image of the object. Initially GI has been demonstrated using quantum approaches such as spontaneous parametric down conversion (SPDC), however, it was later realized that it can equally be conducted with classically correlated twin beams such as chaotic thermal light field [114-116]. The measured correlation functions in both processes provide nearly the same results. Another variants of GI, such as Fourier-plane GI, the measured correlation properties might also represent the Fourier spectrum of the object and computational GI, removes the need for the imaging arm and the spatially resolving CCD detector [117, 118]. Structured coherence or illumination is another means to enhance the performance. Incoherent imaging with x-rays was also reported [119, 120].

Some of the important applications of these two-point correlations are in understanding the beam characteristics on propagation through free space and through turbulent atmosphere [121-130]. Closely related to the objective of speckle reduction is the observation that partially coherent beams are often more useful than their fully coherent counterparts for applications involving propagation through random media such as the turbulent atmosphere. The interest in the reduction of beam distortion on propagation by partial coherence grew dramatically over the years. In initial studies, the focus was on mutual coherence function, intensity fluctuations and transmittance of light beams passing through turbulence [121-129]. Further studies include use of beams emanating from model sources such as GSM beams, and laser beams which are useful for applications in optical communication [130-132]. A turbulence free interferometer has been proposed very recently in which information of correlations can be recovered in the form of second order interference [133].

The benefit of scintillation reduction by partial coherence has led an increasing number of investigations of partially coherent beams specifically for optical communications. Gbur and Wolf [134] theoretically evaluated the spreading of such beams in random media, showing these beams are less sensitive to turbulence; this result was demonstrated experimentally by Dogariu and Amarande [135]. Similar theoretical results were achieved by several researchers in the scientific community globally. A "pseudo- partially coherent beam" was introduced for free-space communication [136]. A number of mathematical tools have been developed for studying partially coherent beams in turbulence [137, 138]. Several properties of partially coherent beams such as coherence length, diffusion, average transmittance etc. after passing through turbulent atmosphere have been explored in detail [139-148]. Two-point correlations have also been widely used in studying beam propagation though tissues and in underwater communications [149-151]. The EM beams have been found to be robust for free-space communication applications [152].

Coherence properties of several kinds of beam shapes and structures have been studied in the past years. These beams include vortex beams, higher order Gaussian, Laguerre Gaussian beams, Bessel beams, etc for propagation and other coherence studies[153-159]. Coherence function studies have been useful in solving the structure determination problems such as involving x-ray diffraction and holography [160, 161]. They have been used to develop experiments for particle/atom trapping [162-164], generating controllable far-field beam profiles etc [165]. Non-linear effects using partially coherent beams were also studied [166]. EGSM beam propagation through lens systems has been investigated in both space-time and space-frequency domain very recently showing the change in DOP of the beam at 2-f and 4-f planes [167-169].

It has been shown that the focusing and diffraction of polychromatic light generally results in spectral changes [170]. In studying the diffraction of partially coherent and broadband light, such rapid changes in spectral density, namely "spectral switch" have been proposed and demonstrated experimentally [171]. Of late, such spectral changes were proposed using polarization and application of spectral switching on data encode, data hiding and in free-space optical communication [172]. Spectral encoding of data holds the promise of low noise fluctuations and low errors due to atmospheric effects during the transmission process. Use of fast switching electro-optic devices such as liquid crystal, Pockel's cell etc. may offer fast switching as the process no longer remains mechanical [173].

More recently, through the quantum theory of optical coherence [174], generation of twin-photons having quantum correlation features has been achieved through the process of SPDC which is having applications in several emerging areas such as quantum computing and quantum information science. Such photons encapsulate spatial entanglement features [175]. It has been demonstrated both theoretically and experimentally that use of partially coherent pump beam having desired degree of coherence may control the spatial coherence properties of the generated biphotons [176]. Additionally, the spatial and spectral profile of the down converted photons depend on the pump coherence features [177]. Since partially coherent beams are more robust compared to fully coherent beams during propagation through atmosphere (free-space), use of partially coherent pump could produce photons with less deleterious effects useful for applications in quantum communication and quantum key distribution [178]. Other applications of optical coherence in quantum domain include quantum computing and fundamental testing [179, 180].

## 10. Conclusion

In this review, we have presented the recent developments in experimental determination of two-point coherence functions in both space-time and in space-frequency domain in the field of electromagnetic beams having partial coherence and partial polarization features. Methods based on first order interference and intensity-intensity correlation were discussed with a focus on their usability. At the end, applications of partial coherence and polarization in several domains of science including tomography, imaging, microscopy, holography, beam propagation through turbulence, spectral switching, optical communication and quantum optics were briefly reviewed. Based on the huge applicability of optical correlations in multiple areas of classical and quantum domain, we believe this field will grow and expand rapidly, and many more interesting results and potential applications will be revealed in the coming years.

**Acknowledgment**

Deepa Joshi gratefully acknowledges financial support received from the Department of Science and Technology (DST), India under Women Scientist (WOSA) scheme.